\documentclass{aastex}
\usepackage{spr-astr-addons}
\usepackage{url}

\RequirePackage{color}

\newcommand{\be}{\begin{equation}}
\newcommand{\ee}{\end{equation}}
\newcommand{\bea}{\begin{eqnarray}}
\newcommand{\eea}{\end{eqnarray}}

\begin{document}

\title{Study of Interplanetary Magnetic Field with Ground State Alignment}
\shorttitle{Study of Interplanetary Magnetic Field with GSA}
\shortauthors{J. Shangguan and H. Yan}

\author{Jinyi Shangguan\altaffilmark{1}} \and \author{Huirong Yan\altaffilmark{2}}
\altaffiltext{1}{KIAA, Peking University, Beijing, 100871, China, shangguan@pku.edu.cn}
\altaffiltext{2}{KIAA, Peking University, Beijing, 100871, China, hryan@pku.edu.cn}

\begin{abstract}
We demonstrate a new way of studying interplanetary magnetic field -- Ground State Alignment (GSA). Instead of sending thousands of space probes, GSA allows magnetic mapping with any ground telescope facilities equipped with spectropolarimeter. The polarization of spectral lines that are pumped by the anisotropic radiation from the Sun is influenced by the magnetic realignment, which happens for magnetic field ($<1G$). As a result, the linear polarization becomes an excellent tracer of the embedded magnetic field. The method is illustrated by our synthetic observations of the Jupiter's Io and comet Halley. Polarization at each point was constructed according to the local magnetic field detected by spacecrafts. Both spatial and temporal variations of turbulent magnetic field can be traced with this technique as well. The influence of magnetic field on the polarization of scattered light is discussed in detail. For remote regions like the IBEX ribbons discovered at the boundary of interstellar medium, GSA provides a unique diagnostics of magnetic field.
\end{abstract}


\section{Introduction}

Magnetic fields are ubiquitous and play extremely important roles in many astrophysical circumstances, e.g., the interstellar medium, intergalactic medium, and quasars. In the mean time, there are only a few techniques for magnetic field studies available. Each technique is sensitive to magnetic field in a particular environment and has its own limitations. Therefore, even the direction of a magnetic field obtained for the same region of sky with different techniques differs substantially. The synergy use of different techniques provides a possibility of magnetic field tomography.(see \citealt{Yan2006} for details)

Interplanetary magnetic field is turbulent and it plays important role for many processes in the interplanetary diffuse media, e.g., star formation, accretion and the cosmic ray propagation and acceleration, etc. The recently launched NASA's satellite Interstellar Boundary Explorer (IBEX) \citep{McComas} discovered a bright narrow "ribbon" of energetic neutral atoms (ENA) emission nearly across the entire celestial sphere, which is unpredicted by any earlier models or theories. It has been shown that this ribbon is largely stable, with small time variations, which implies a sustained local magnetic structure interacting with the original solar magnetic field (\citealt{Funsten} and \citealt{Schwadron}). Ground State Alignment (henceforth GSA) we discuss in this paper provides a unique chance to measure the fine scale structure of 3D magnetic field.

GSA is a new promising way of studying magnetic fields in radiation-dominated environments. In fact, this diagnostic is far more sensitive than those based on the Zeeman effect for the magnetic field in the diffuse medium. The basic idea of GSA is simple: consider atoms irradiated by a nearby star. Anisotropic radiation pumps the atoms differentlly from different magnetic sublevels, resulting in over- or underpopulation of the magnetic sublevels on the ground state. Then the absorbed and scattered light from the aligned atoms will get polarized, depending on the diection of magnetic field. Therefore GSA is an effective way to detect the 3D orientation of magnetic field.

It is worthwhile mentioning that optical pumping was first proposed by \citet{Kastler1950}, who won the Nobel prize in 1966 for the work. The optical pumping process can induce asymmetry on the ground state. Later experiments (\citealt{Brossel1952} and \citealt{Hawkins1953}) confirmed this idea. Optical pumping has a lot of applications on laboratory physics, however, the application to astrophysics up to now is only Hanle effect in stellar atmosphere and masers (\citealt{Litvak1966}, \citealt{Perkins1966}, \citealt{Mies1974} and \citealt{Rausch1996}). GSA was noticed and invoked in the interstellar case by \citet{Varshalovich1968} in the case of hyperfine structure of the ground state. A more rigorous study of using GSA to diagnose weak magnetic fields in diffuse media was conducted in \citet{Landolfi}. But, the condition they employed is too idealized to be used in any real astrophysical environments. Yan and Lazarian (2006, 2007 and 2008) provided calculations of absorption and emission lines with fine or hyperfine structure in a more general condition, which opened the avenue for the application of GSA to astronomy (See \citealt{Yan2012} for more discussion.)

Here we emphasize that the GSA we discuss in this paper differs from the Hanle effect employed in the solar magnetism. The magnitude of magnetic field detectable through GSA is in the sub-gauss regime, the regime of magnetic field in diffuse medium. In this case, the effect of magnetic field is mainly on the ground state, where magnetic mixing happens, as long as the Larmor frequency is higher than the excitation rate of the atom's ground level. Compared with this, however, the Hanle effect takes place in the higher magnetic field regime. Hanle effect requires the magnetic field to influence the upper level of atoms, thus Larmor frequency should be high enough to be comparable with the Einstein coefficient of the atom's upper level (see \citealt{Trujillo}). Moreover, most of the diffuse media, for example interplanetary media, have the magnetic field in the regime of GSA, so we consider only this effect.

In this paper, we apply the GSA to synthetic observations of Jupiter's Io and comet Halley. We use the magnetic field data of the two objects, detected by space probes before.\footnote{Galileo spacecraft was sent to Jupiter for an 8-year mission in Jovian system. We downloaded the data of Jupiter's magnetic field from NASA PDS-PPI (see \citealt{Kivelson}). Vega spacecrafts (vega1 and vega2) were sent to Venus and to comet Halley. This mission was implemented by the USSR with international participation. The main spacecraft encounter with Halley was in March 6 and 9, 1986. Data were obtained during both the encounter and cruise phases of the mission. See \url{ftp://nssdcftp.gsfc.nasa.gov/spacecraftdata/vega/mag/}} Namely, we synthetically observe the sodium emission D lines of the two sources, partially motivated by the fact that there is abundant sodium in the wake of Io and the body of Halley. The alignment of sodium was calculated in \citet{Yan2007}. 

Undoubtedly, compared to the vast interplanetary space that magnetic field prevails in, the information we can get from spacecrafts is far less than adequate. With this study, we try to demonstrate GSA as an easy but powerful magnetic diagnostic. It even has higher space resolution than that of the direct detection in space. It enables a tomography of interplanetary magnetic field without sending thousands of space probes.

In what follows, we first introduce the basic idea of GSA and how the magnetic field is involved in this process in \S2. Then we present the synthetic observation of the magnetic fields traced by Io and Halley in \S3 and discuss the influence of the magnetic field on the polarization of the scattering light in \S4. In \S5,  the feasibility of real observation is discussed. The discussion and the summary are provided in \S6 and \S7, respectively.

\section{BRIEF INTRODUCTION OF GSA}

\begin{figure}[!ht]
\includegraphics[
  width=0.4\textwidth,
  height=0.3\textheight]{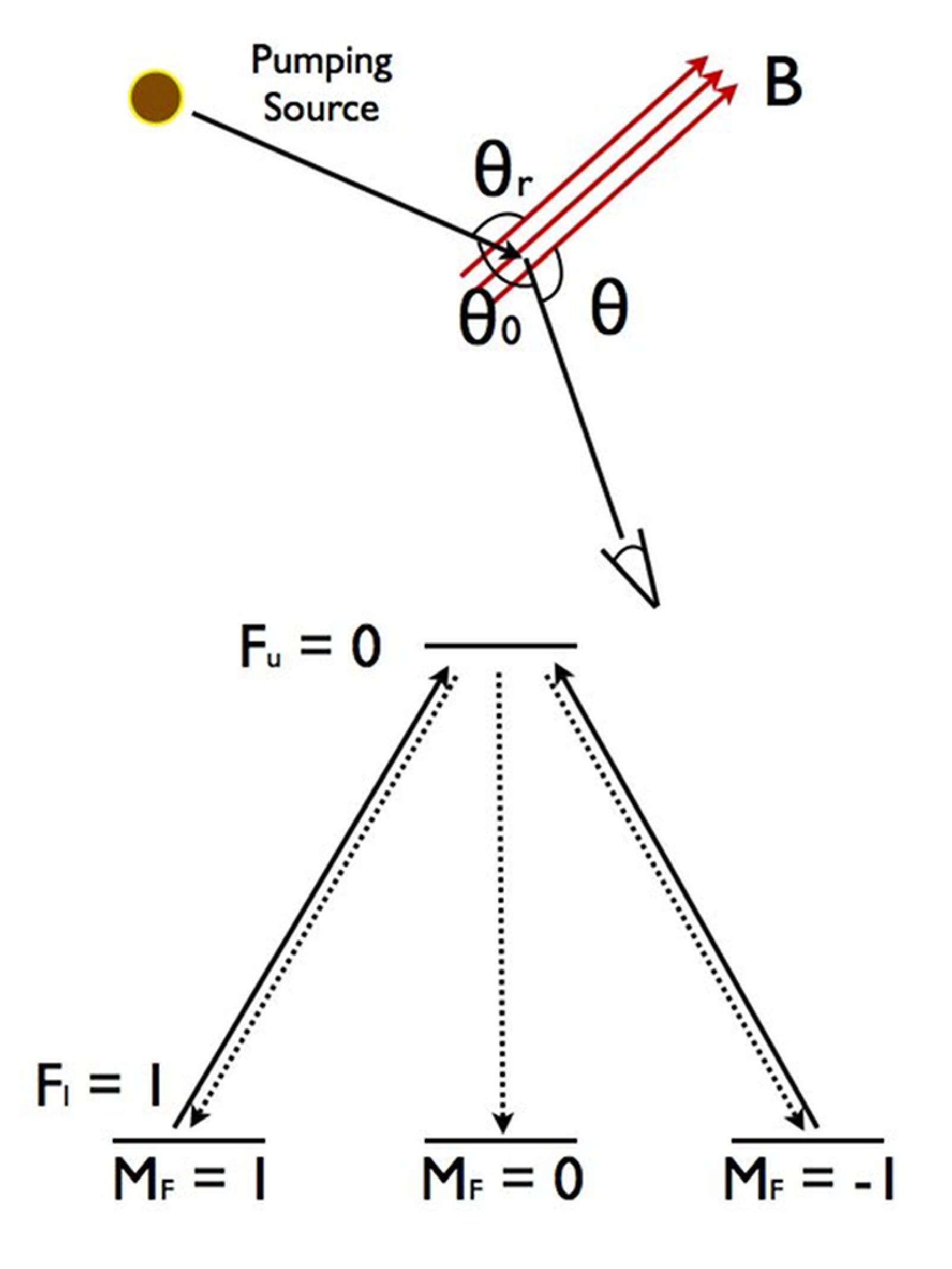}
\caption{{\it Upper} : Typical environment where GSA happens. $\theta_r$ is the angle between incoming radiation and magnetic field. $\theta$ is the angle between line of sight and magnetic field. $\theta_0$ is the angle between incoming radiation and line of sight, which is also called scattering angle. {\it Lower}: Toy model to illustrate the atoms aligned on $M_F=0$ sublevel by anisotropic light. From \citet{Yan2008}.}
\label{nzplane}
\end{figure}

Atoms'\footnote{Since the alignment of atom and ion are generally the same, for simplicity we use atom to refer both of them in this paper.} angular momenta can be aligned by the anisotropic radiation flux. Owing to this, the scattered light from these aligned atoms is polarized. Magnetic field, if exists, will modify the alignment by mixing magnetic sublevels with different angular momentum if the Larmor frequency is higher than the rate of the excitation from the ground state. The information of the magnetic field, therefore, is imprinted in the polarization of light. The term GSA refers to both radiative alignment and magnetic realignment (see \citealt{Yan2012} for details). The typical geometry of GSA is shown in Fig.\ref{nzplane} {\it upper}.

We discuss a toy model to illustrate further the physics of GSA (see Fig.\ref{nzplane} {\it lower}). Let us consider an atom with total angular momenta $F_l=1$, $F_u=0$ in its ground state and the upper state, respectively. We denote ``$M_F$" as the projection of the angular momentum to the direction of the incident radiation. On the ground level, there are three sublevels, namely $M_F=-1$, 0 and 1. Since a photon beam consists of only left and right circularly polarized photons, the transitions from the ground state to the upper state happens only from the $M_F=-1$ and 1 sublevels of the ground state to the upper state ($M_F=0$). Meanwhile, the atoms decay with equal probabilities from the upper state to all three sublevels of the ground state. As the result, the atoms in the ground state are concentrated on the $M_F=0$ sublevel. Therefore the optical properties of the medium are changed and the light scattered by this medium will be polarized. Any atoms  with quantum angular momentum $\geq1$ in a long-lived state can be aligned. Two main factors affect the alignment. One is collision which thermalize the distribution among the sublevels on the ground state and therefore destroy the alignment. Another is the magnetic field as discussed at the beginning of this section.

For the optically thin case, the linear polarization degree, the position angle and the line ratio $I_{D2}/I_{D1}$ are
\be
p=\sqrt{Q^2+U^2}/I=\sqrt{\epsilon_2^2+\epsilon_1^2}/\epsilon_0,
\label{pchi1}
\ee
\be
\chi=\frac{1}{2}\tan^{-1}(U/Q)=\frac{1}{2}\tan^{-1}(\epsilon_2/\epsilon_1),
\label{pchi2}
\ee
\be
\frac{I_{D2}}{I_{D1}}=\epsilon_0(D2)/\epsilon_0(D1).
\label{Lr}
\ee
where $\epsilon_i$ ($i=1,2,3$) are emission coefficients for the Stokes parameters. The expression of $\epsilon_i$ can be found in \cite{Yan2007} as the solution of statistical equilibrium equations for the hyperfine transitions on both upper and lower levels.

\section{SYNTHETIC OBSERVATION}

\begin{figure}[!ht]
\centering
  \includegraphics[
  width=0.35\textwidth,
  height=0.25\textheight]{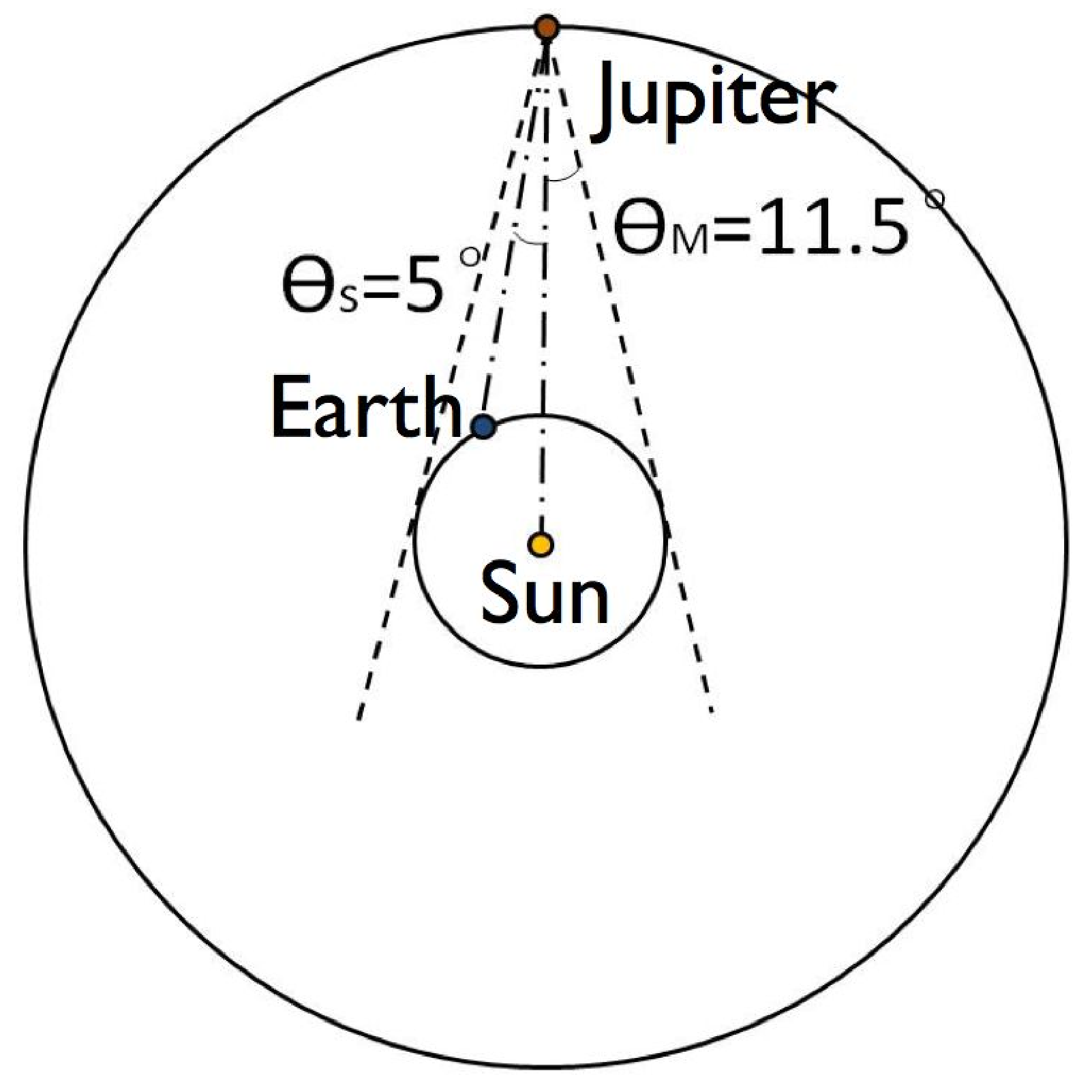}
  \caption{The geometry of the Sun, Earth and Jupiter. $\theta_M$ refers to the maximum Jupiter phase viewed from the Earth. However, in our synthetic observation we set the phase ${\theta_s=5^\circ}$.}
  \label{obsergeometry1}
\end{figure}

\begin{figure}[!ht]
\centering
  \includegraphics[
  width=0.4\textwidth,
  height=0.25\textheight]{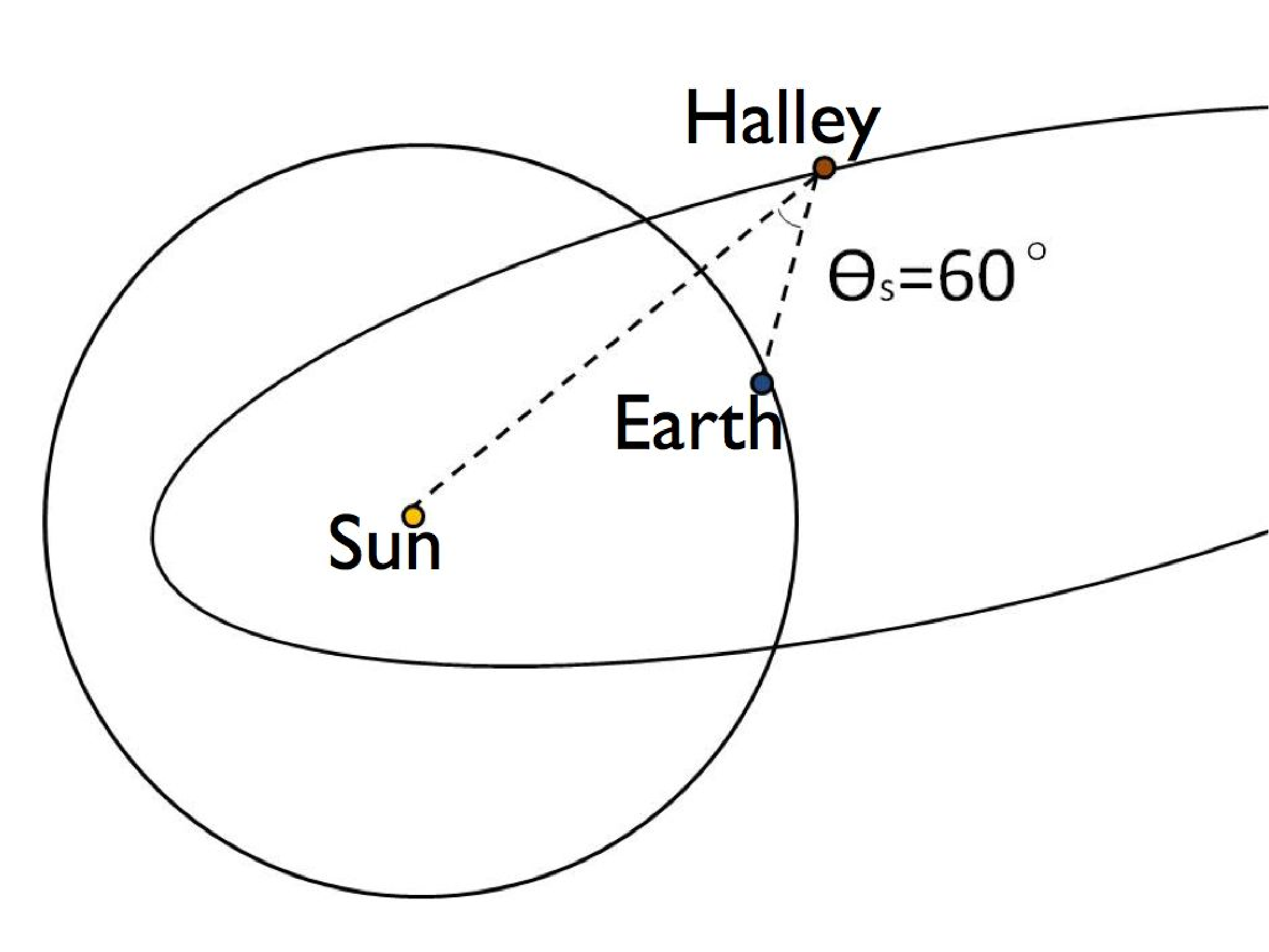}
  \caption{The geometry of the Sun, Earth and Halley. We set the phase ${\theta_s=60^\circ}$.}
  \label{obsergeometry2}
\end{figure}

\begin{figure}[!ht]
  \includegraphics[
  width=0.45\textwidth,
  height=0.3\textheight]{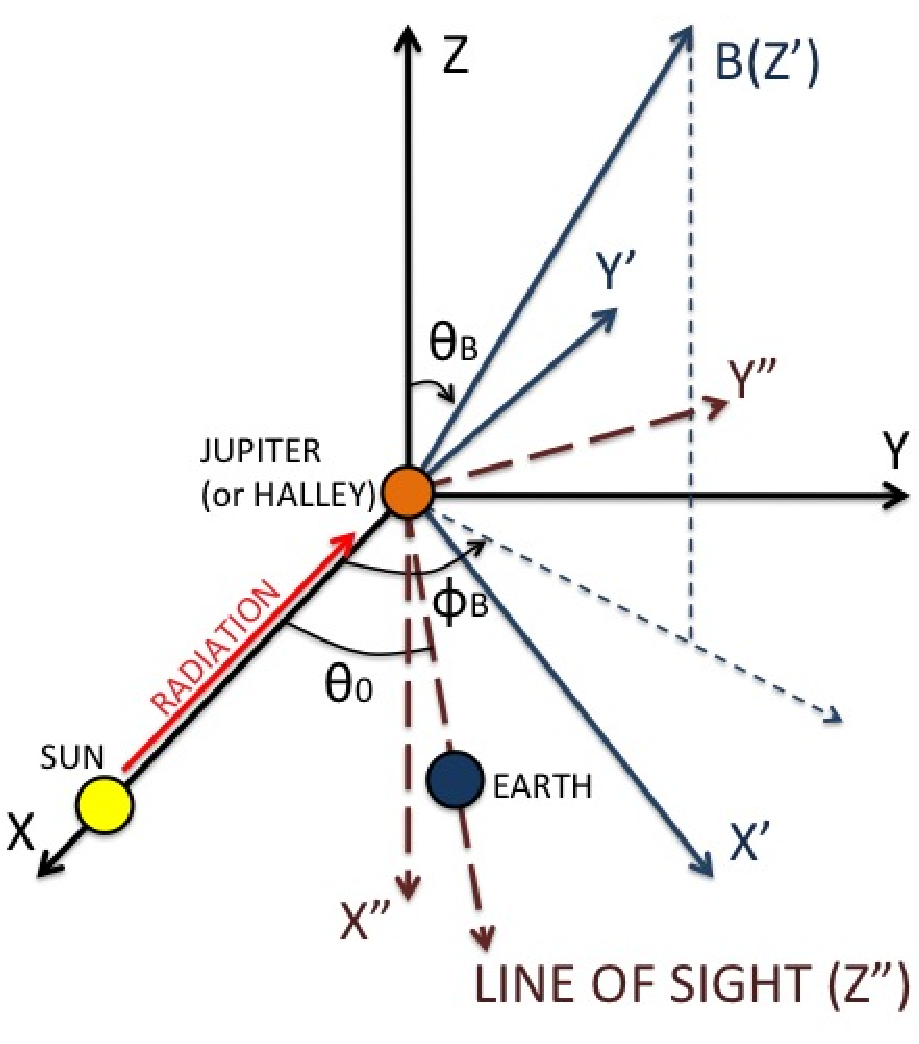}
  \caption{The magnetic field ($\theta_B$, $\phi_B$) was measured by the spacecrafts in the xyz reference frame (the detection frame), where x-axis points from the Jupiter or Halley to the Sun, z-axis is perpendicular to the ecliptic plane of the solar system and y-axis completes the right handed set. Scattering angle, $\theta_0$ is fixed. The theoretical calculation is done in x'y'z' reference frame (the theoretical frame). Therefore, two successive rotations are necessary for frame transformation. First we rotate xyz-system about z-axis by $\phi_B$ to reach an intermediate frame. Then we rotate the new system about the new y-axis by $\theta_B$ to get x'y'z'-system. In x'y'z'-system, the direction of pumping radiation (from the Sun) and line of sight are in  ($\theta_r,\phi_r$) and ($\theta,\phi$), respectively. Since scattering angle is fixed, ($\theta_r,\phi_r$) and ($\theta,\phi$) are confined, and there are still two physical dimensions of freedom. To reach x"y"z" reference frame (the observational frame) from the detection frame, we can first rotate xyz-system about z-axis by $\theta_0$ to get an intermediate frame and  then rotate the new frame about the new y-axis by $90^\circ$.}
  \label{rotate_dethe}
\end{figure}

The geometry of the Sun-Earth-Io (or Halley) scattering system is shown in Fig.\ref{obsergeometry1} and \ref{obsergeometry2}. Viewing from the Earth, the maximum phase of Jupiter is about $11.5^\circ$. Meanwhile, the phase of Halley can be any value in principal from $0^\circ$ to $90^\circ$.  We choose the phase of Io to be $\theta_s=5^\circ$ and that of Halley to be $\theta_s=60^\circ$ in our synthetic observations. The coordinate systems employed in magnetic field detection (the detection frame, xyz-system in Fig.\ref{rotate_dethe}) by the spacecraft missions are almost the same. The details of the coordinate systems can be found in Appendix.

\begin{figure}[!ht]
  \includegraphics[
  width=0.45\textwidth,
  height=0.3\textheight]{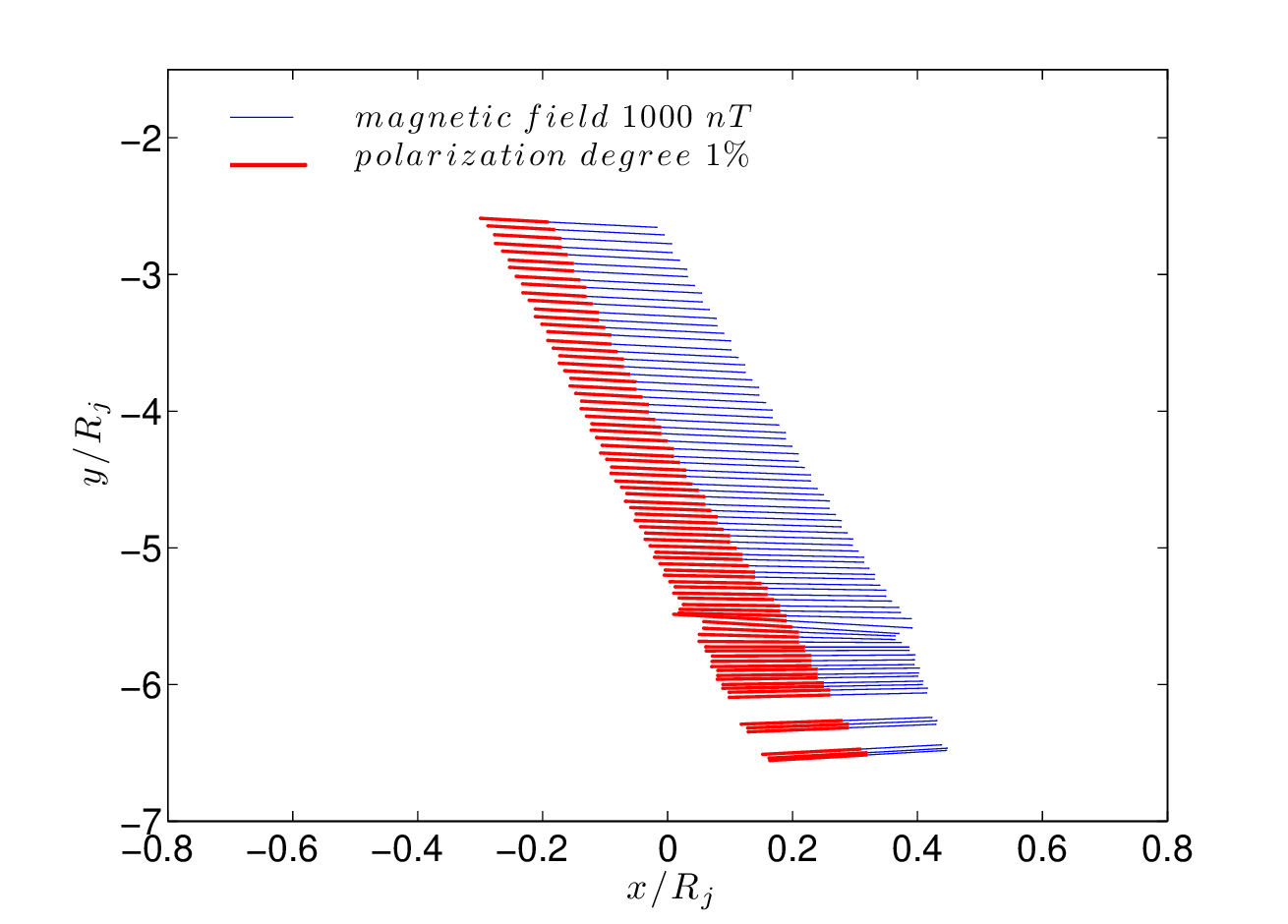}
  \caption{The synthetic observation result of Io shown in celestial coordinate (x-y plane of the observational frame), where the red thick line represents the polarization direction and the blue thin line represents the direction of magnetic field. The x-axis is normal to the ecliptic plane of the solar system. The unit of the coordinates is Rj=71492 km, the equatorial radius of Jupiter. The scale of each physical quantity is demonstrated in the northwest.}
  \label{syn Io traj}
\end{figure}

\begin{figure}[!ht] 
  \includegraphics[
  width=0.45\textwidth,
  height=0.3\textheight]{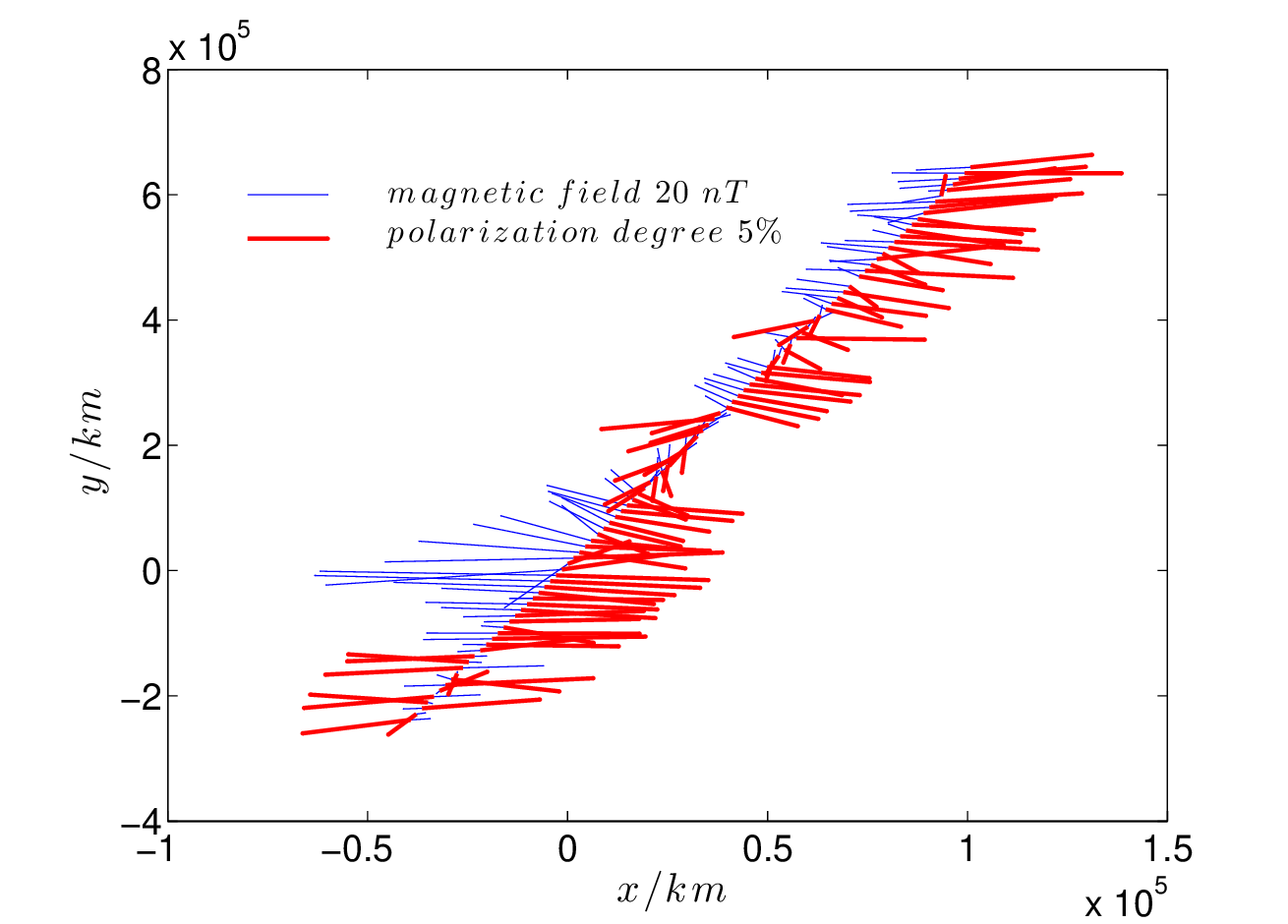}
  \caption{The Magnetic field and polarization along the trajectory of vega1 encountering the comet Halley. The blue thin lines represent magnetic field and the red thick lines represent the polarization vectors. The x-y plane is the plane of celestial coordinates. The x-axis has the direction perpendicular to the ecliptic plane of the solar system.}
  \label{syn Com traj}
\end{figure}

In order to calculate the emission coefficients in the synthetic observations, we need the  following input parameters: the directions of the Sun and the Earth, ($\theta_r,\phi_r$) and ($\theta,\phi$) respectively, in the coordinate system for theoretical calculation (or the theoretical frame, x'y'z'-system in Fig.\ref{rotate_dethe}) \footnote{In this coordinate system, direction of magnetic field is chosen to be the z-axis.}. The conversion of the coordinate system from the detection frame (xyz) to the theoretical frame (x'y'z') is described in Fig.\ref{rotate_dethe} as two successive rotations of the coordinate system. In each specific observation, the scattering angle is known, e.g. $\theta_0=5^\circ$ for Io and $\theta_0=60^\circ$ for Halley in our synthetic observation. Since $\theta_0$ is fixed, there are only two physical dimensions of freedom (the direction of the magnetic field) for the scattering system as shown in Fig.\ref{nzplane} {\it upper}. When polarization is calculated, the results (Figs.\ref{syn Io traj} and \ref{syn Com traj}) are shown in the observational coordinate system (or the observational frame, x"y"z"-system in Fig.\ref{rotate_dethe}) where z"-axis points to the Earth and x"-axis is perpendicular to the ecliptic plane of the solar system. The rotation from the detection frame (xyz) to the observational frame (x"y"z") follows the routine demonstrated in Fig.\ref{rotate_dethe}.

\begin{figure}[!ht]
   \includegraphics[
  width=0.45\textwidth,
  height=0.3\textheight]{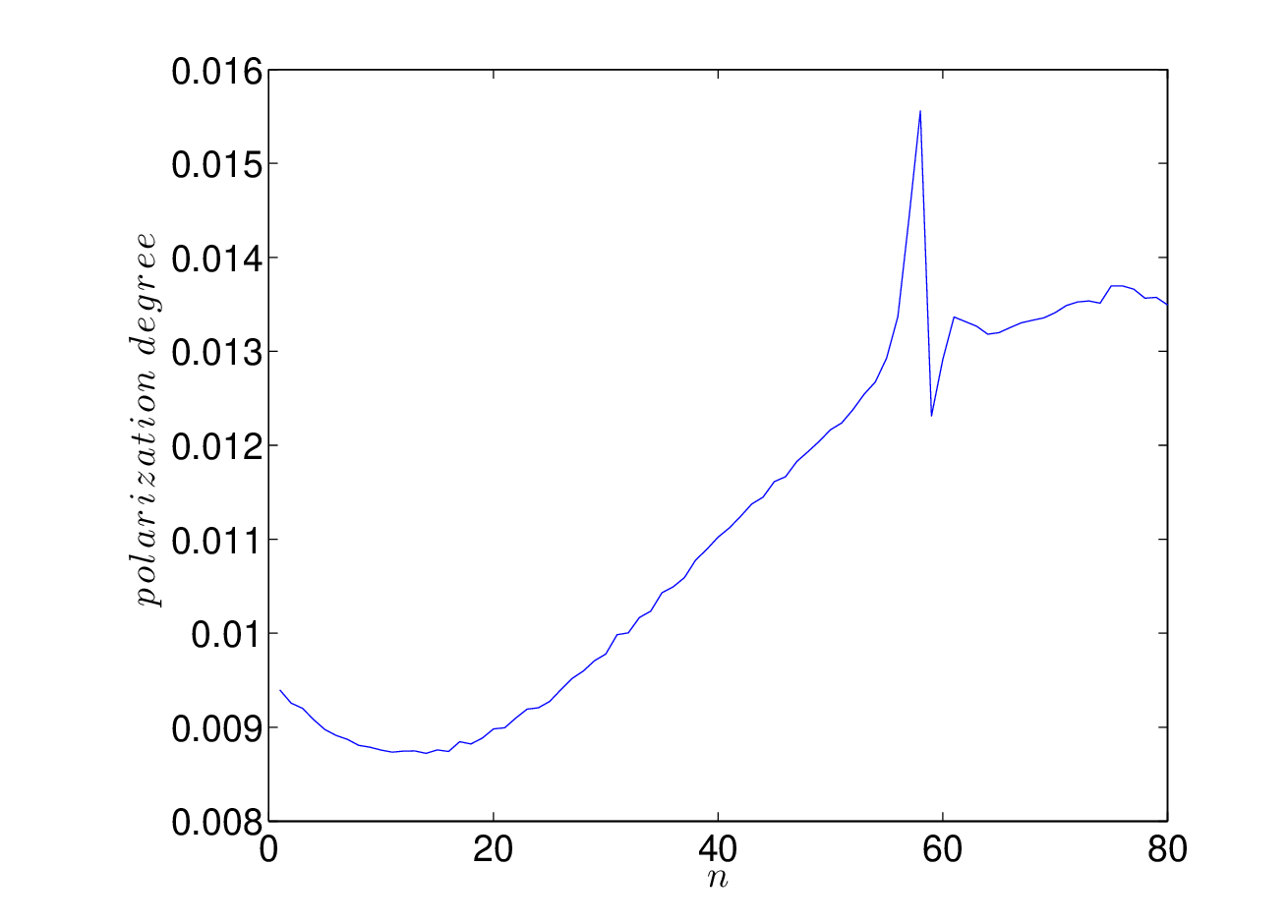}
  \caption{The polarization degree of Io for each magnetic field data. {\it n} of the x-axis is the sequence number of the data.}
  \label{syn Io pol}
\end{figure}

\begin{figure}[!ht]
  \includegraphics[
  width=0.45\textwidth,
  height=0.3\textheight]{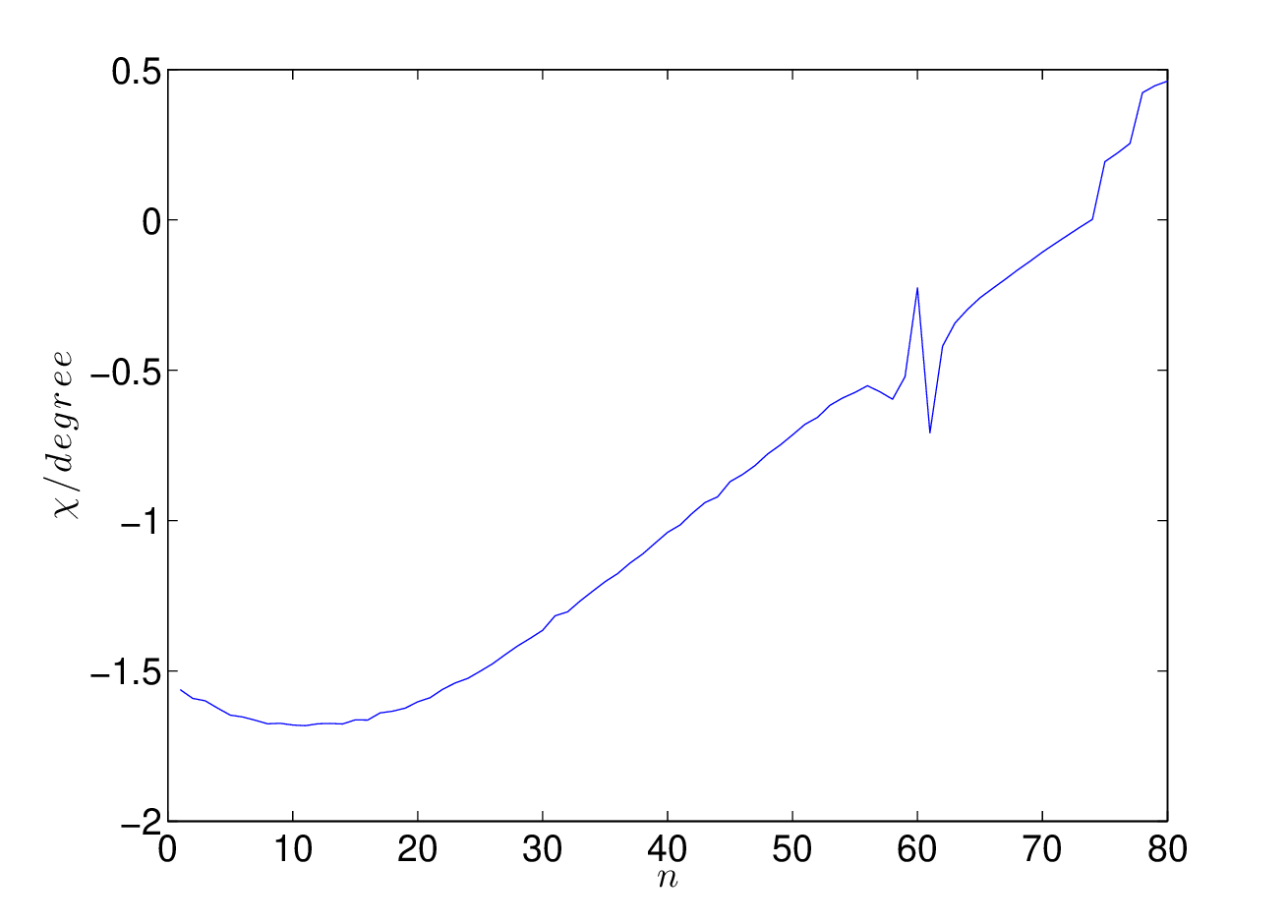}
  \caption{$\chi$ is the position angle of Io for each magnetic field data. {\it n} of the x-axis is the sequence number of the data.}
  \label{syn Io chi}
\end{figure}

\begin{figure}[!ht]
  \includegraphics[
  width=0.45\textwidth,
  height=0.3\textheight]{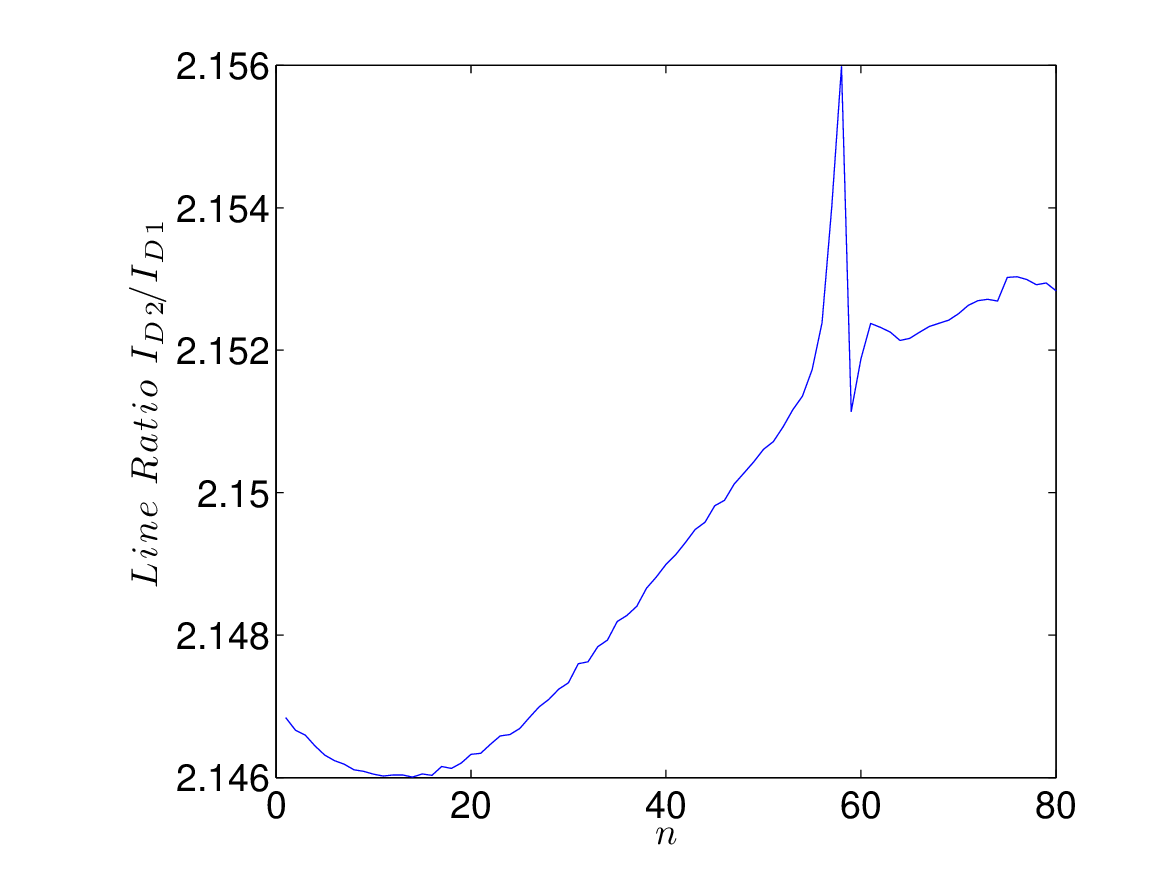}
  \caption{The line ratio $I_{D2}/I_{D1}$ of Io for each magnetic field data. {\it n} of the x-axis is the sequence number of the data.}
  \label{syn Io lr}
\end{figure}

\begin{figure}[!ht] 
  \includegraphics[
  width=0.45\textwidth,
  height=0.3\textheight]{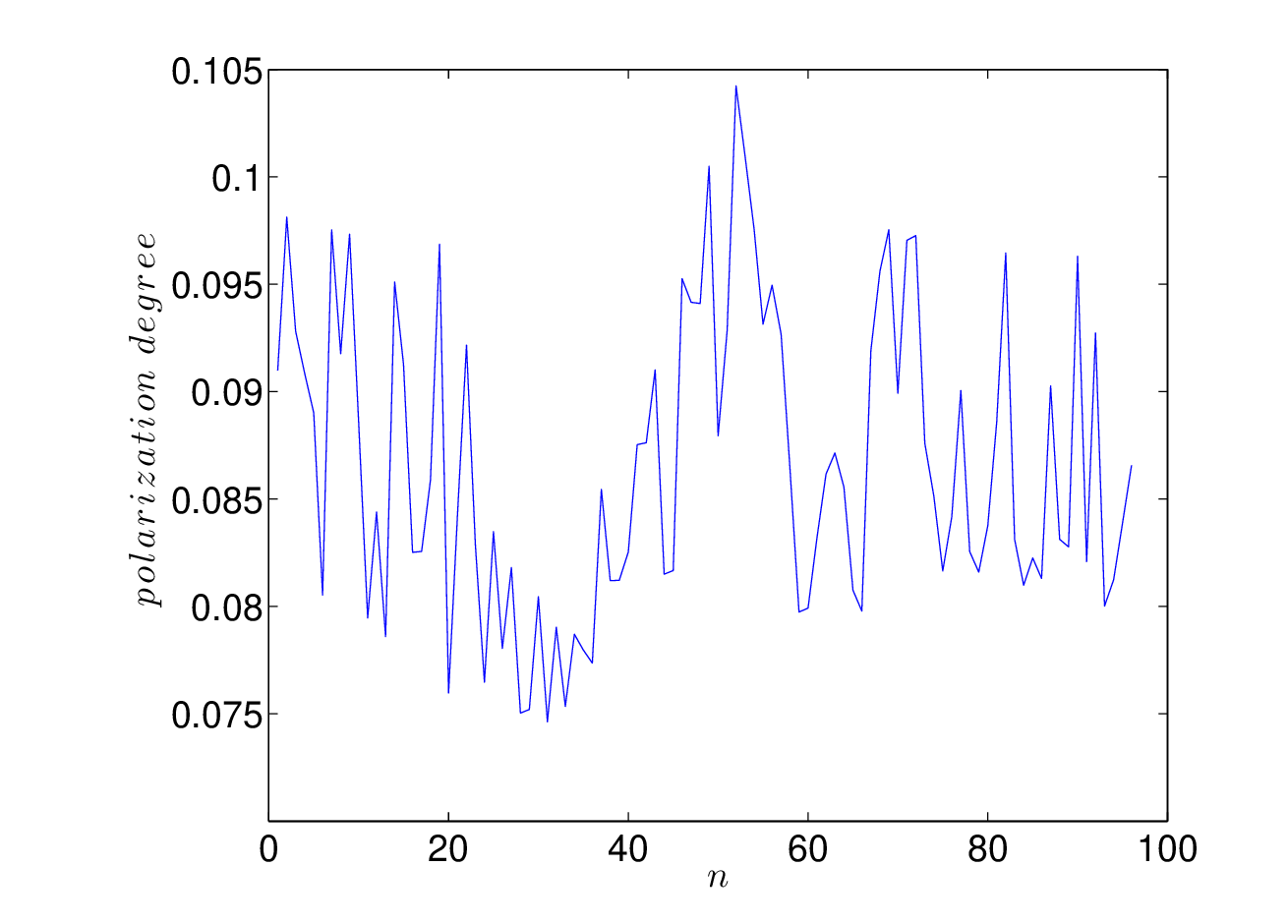}
  \caption{The polarization degree of Halley for each field data. {\it n} of the x-axis is the sequence number of the data.}
  \label{syn Com pol}
\end{figure}

\begin{figure}[!ht] 
  \includegraphics[
  width=0.45\textwidth,
  height=0.3\textheight]{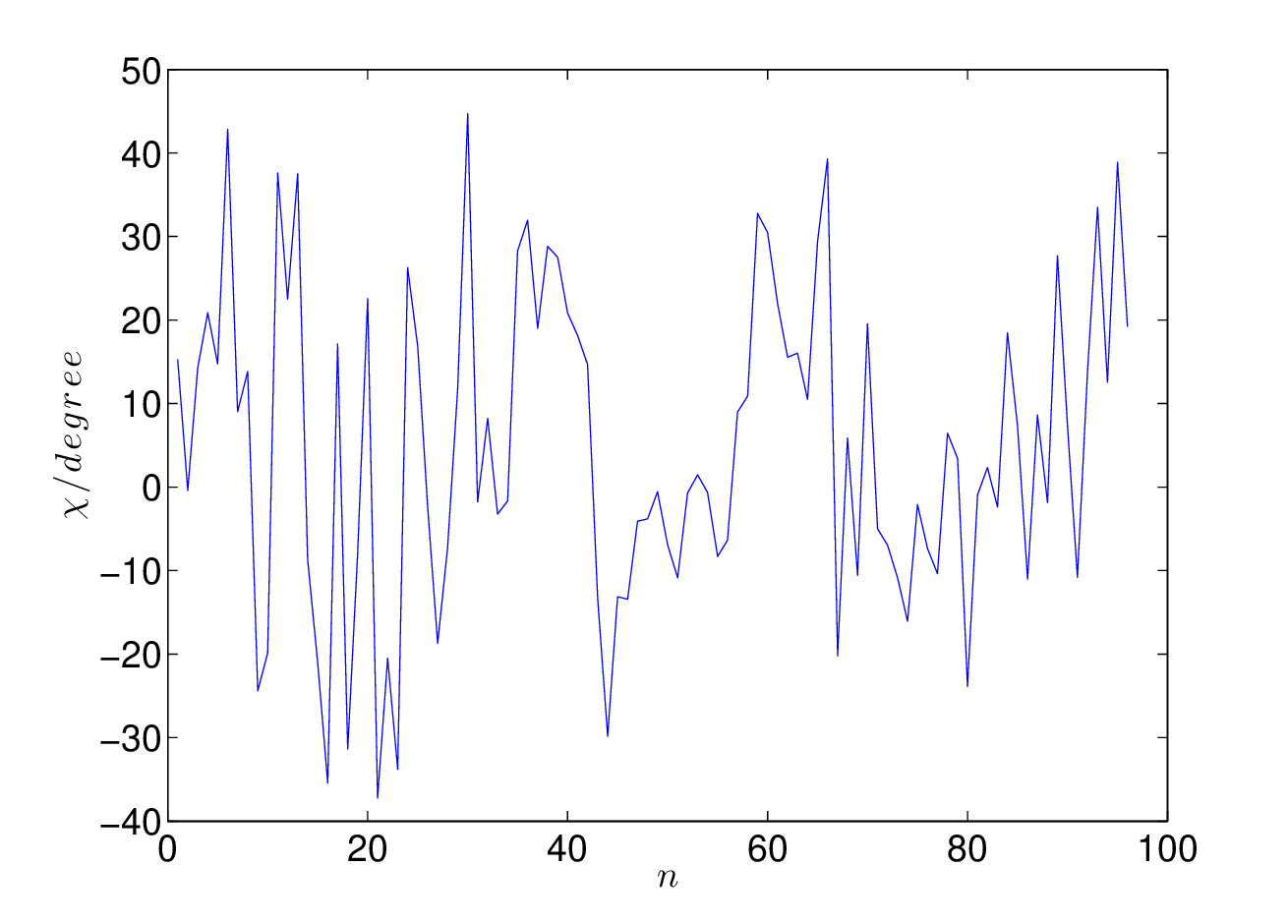}  
  \caption{$\chi$ is the position angle of Halley for each field data. {\it n} of the x-axis is the sequence number of the data.}
  \label{syn Com chi}
\end{figure}

\begin{figure}[!ht] 
  \includegraphics[
  width=0.45\textwidth,
  height=0.3\textheight]{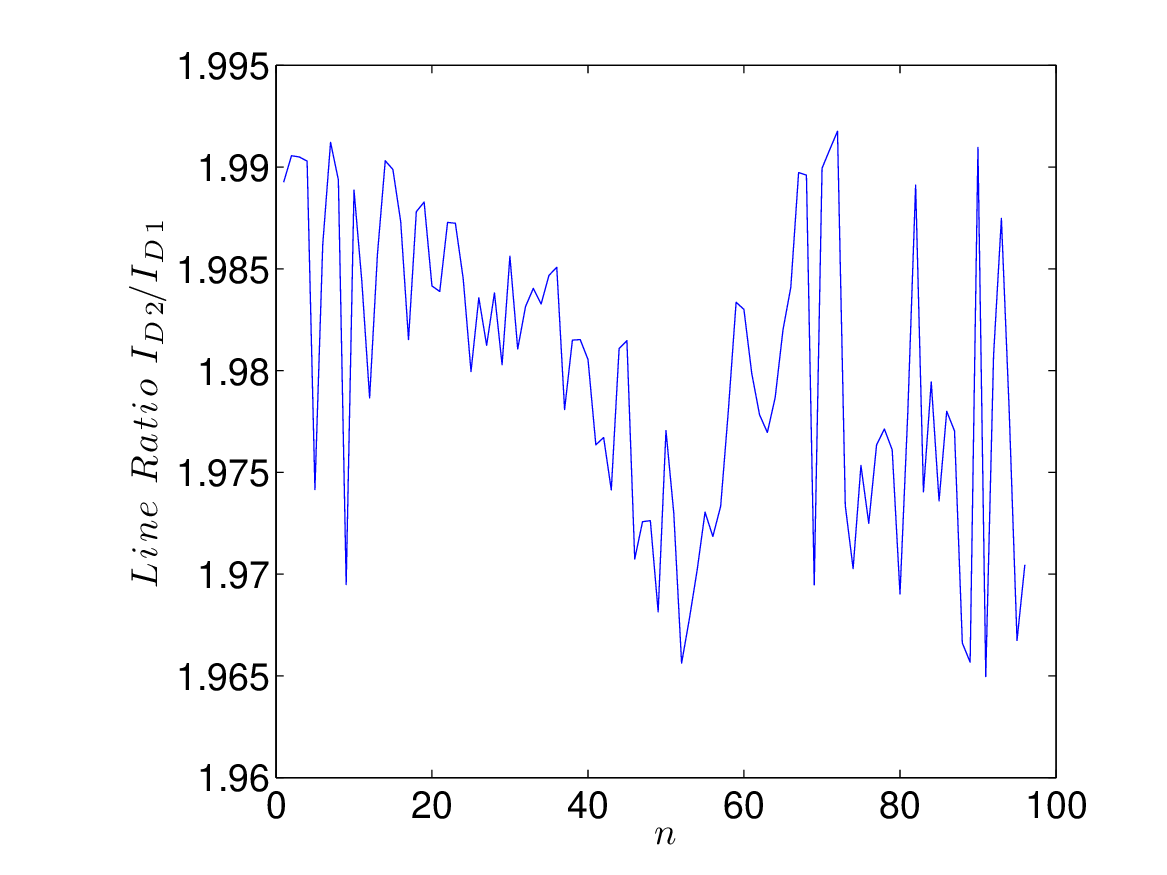}  
  \caption{The line ratio $I_{D2}/I_{D1}$ of Halley for each field data. {\it n} of the x-axis is the sequence number of the data.}
  \label{syn Com lr}
\end{figure}

Since spacecrafts can only get the localized information of magnetic field on its orbit, we follow the trajectories of spacecrafts in our synthetic observation (see Fig.\ref{syn Io traj} and Fig.\ref{syn Com traj}). 
Only a small portion of the spacecraft trajectory has overlap with the Io's orbit around the Jupiter, as shown by Fig.\ref{syn Io traj}. To ensure the clarity of the figures, we use lines instead of arrows to represent the vectors. The blue thin lines show the direction of magnetic field, detected by the spacecraft, and the red thick lines illustrate the?corresponding polarization at each location.

The average polarization degree of Io is around $1.12\%$, and the highest polarization degree reaches $1.56\%$ (see Fig.\ref{syn Io pol}). Fig.\ref{syn Io chi} shows the variation of the position angle $\chi$, the angle between the polarization vector and the vector of magnetic field projected to the celestial coordinate plane. $\chi$ is very small ($-1.7^\circ \sim 0.5^\circ$), since the scattering angle is small in the case of Io (see \S4 for detailed discussions). And the line ratio of D2 line over D1 line for Halley is shown in Fig.\ref{syn Io lr}. As for Halley, the mean polarization degree is $8.66\%$ and  the maximum is $10.4\%$ (see Fig.\ref{syn Com pol}). Fig.\ref{syn Com chi} and \ref{syn Com lr} show the variation of position angle and line ratio $I_{D2}/I_{D1}$, respectively. Since comets are cruising around the interplanetary space, it is a cost effective way to detect the interplanetary magnetic field by tracing the Na emission from comets.

\section{INFLUENCE OF MAGNETIC FIELD ON POLARIZATION}

\begin{figure}[!ht]
  \includegraphics[
  width=0.45\textwidth,
  height=0.3\textheight]{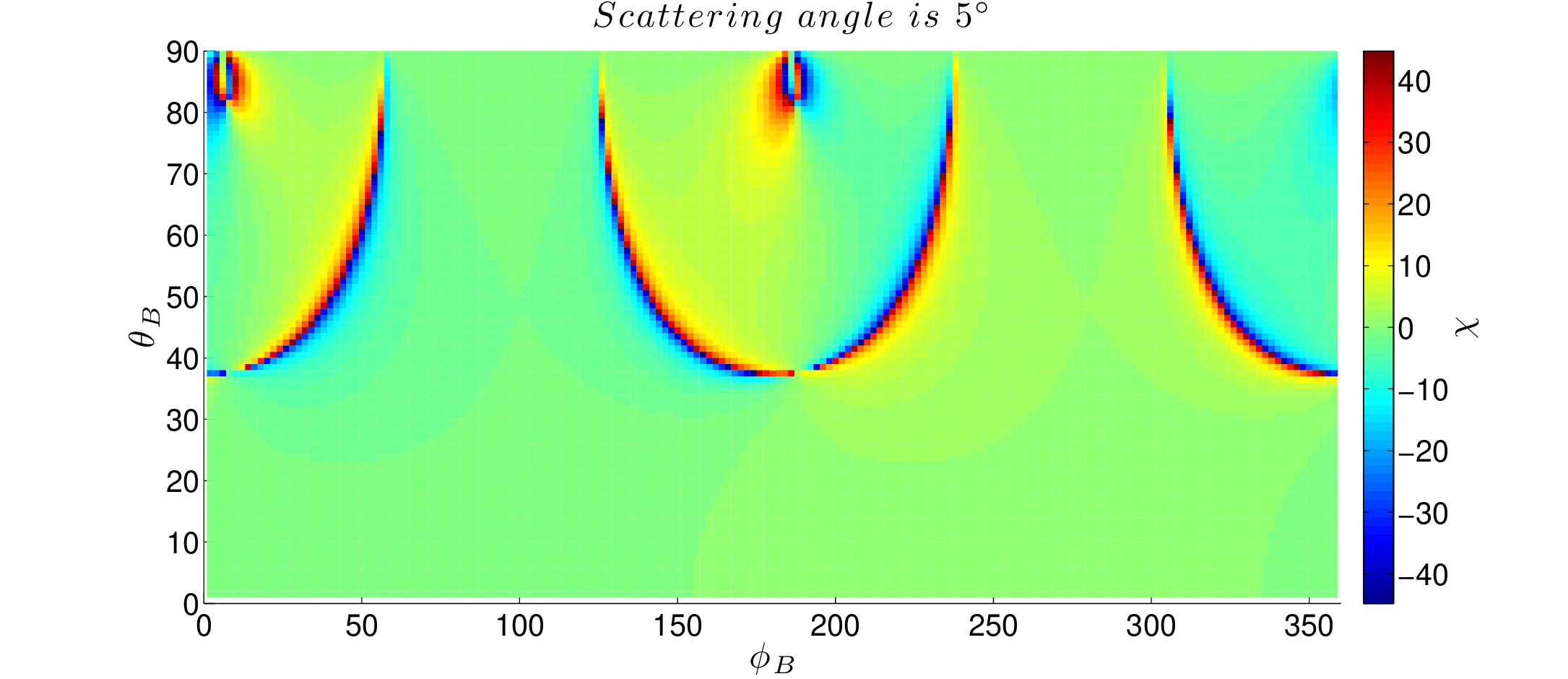}
  \caption{The contour map of position angle, $\chi$, as a function of the direction of magnetic field. The scattering angle is set to be $5^{\circ}$. $\theta_{B}$ and $\phi_{B}$ are defined in Fig.\ref{rotate_dethe}.}
  \label{chi_analysis_5}
\end{figure}

\begin{figure}[!ht] 
  \includegraphics[
  width=0.45\textwidth,
  height=0.3\textheight]{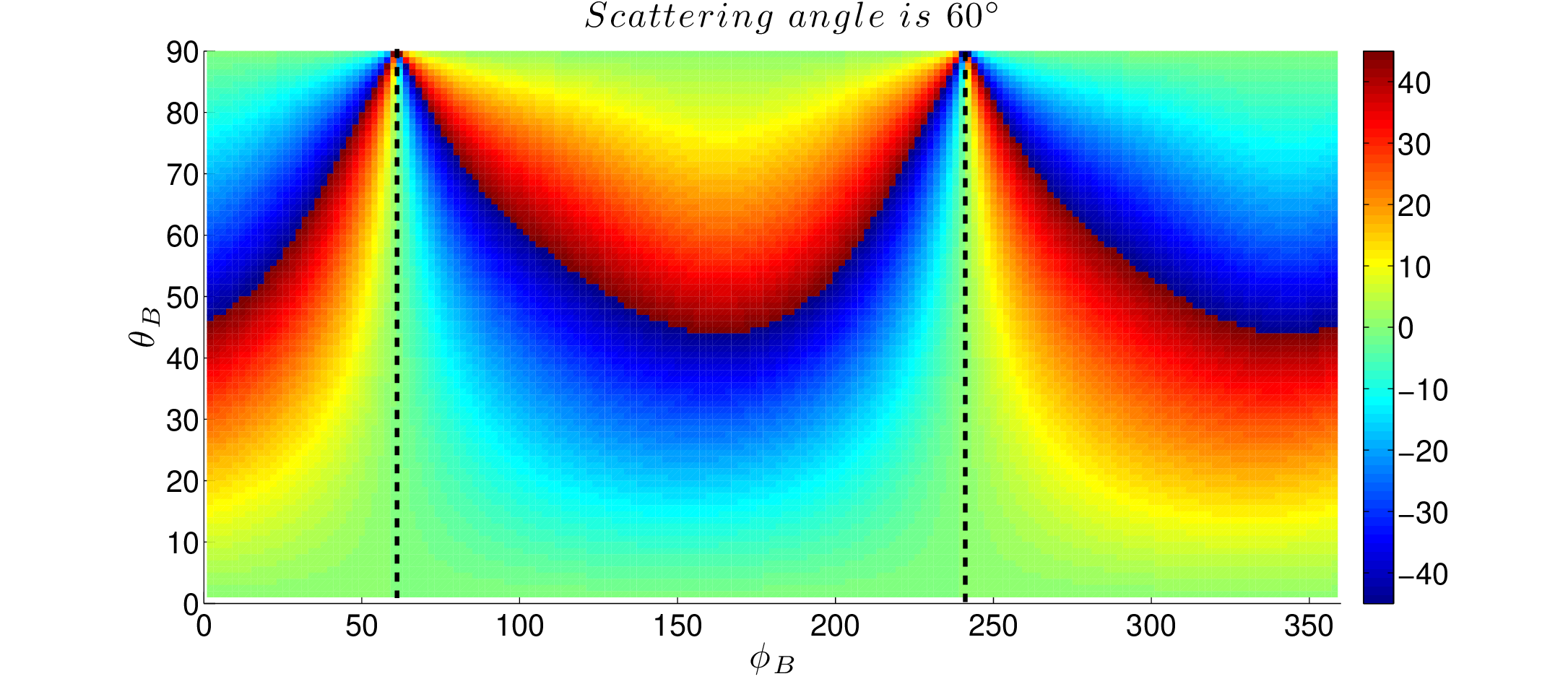}
  \caption{The contour map of position angle, $\chi$, as a function of the direction of magnetic field. The scattering angle is set to be $60^{\circ}$. $\theta_{B}$ and $\phi_{B}$are defined in Fig.\ref{rotate_dethe}. The dashed lines are two quasi-symmetry axes corresponding to $\phi_{B}=60^\circ$ and $240^\circ$, respectively.}
  \label{chi_analysis_60}
\end{figure}

\begin{figure}[!ht] 
  \includegraphics[
  width=0.45\textwidth,
  height=0.3\textheight]{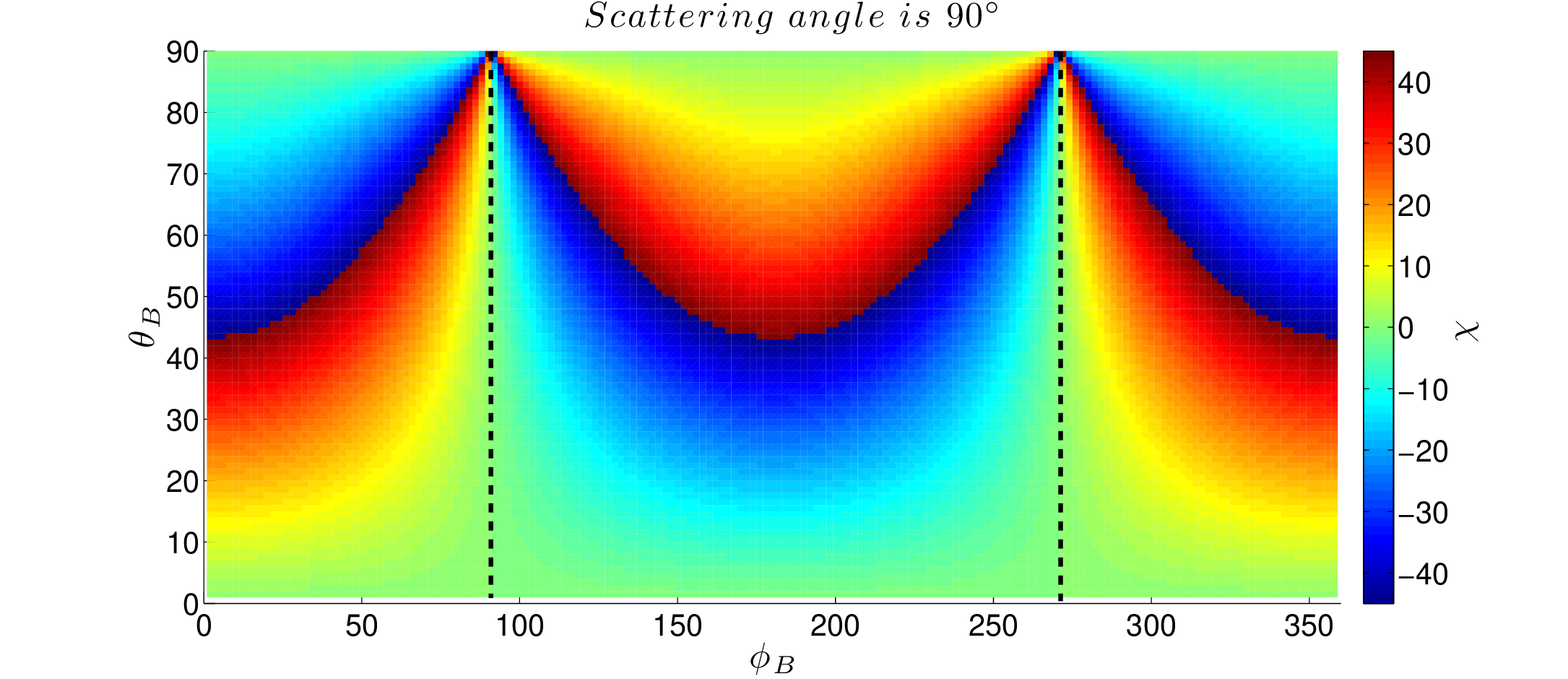}
  \caption{The contour map of position angle, $\chi$, as a function of the direction of magnetic field. The scattering angle is set to be $90^{\circ}$. $\theta_{B}$ and $\phi_{B}$ are defined in Fig.\ref{rotate_dethe}. The dashed lines are two symmetry axes corresponding to $\phi_{B}=90^\circ$ and $270^\circ$.}
  \label{chi_analysis_90}
\end{figure}

To analyze the influence of magnetic field, we provide several contour maps of position angle as a function of magnetic field direction for fixed scattering angle. The geometry of the maps below is defined in Fig.\ref{rotate_dethe}. The xyz-system is employed here. The scattering plane is fixed on x-y plane and the direction of magnetic field described by ($\theta_B, \phi_B$)\footnote{For the sake of symmetry, we only consider the domain with $0 \leq\theta_B\leq 90^\circ$}. Fig.\ref{chi_analysis_5} shows the position angle ($\chi$) map where scattering angle is set to be $5^\circ$ and the direction of magnetic field varies. Fig.\ref{chi_analysis_60} and Fig.\ref{chi_analysis_90} are also $\chi$ maps with scattering angle $\theta_0$ fixed at $60^\circ$ and $90^\circ$ respectively. There are several obvious parabolic curves in each maps, on which the polarization degree reduces to zero and $\chi$ jumps from $45^\circ$ to $-45^\circ$ (or inversely). The shape of the curves and the color distributions in Fig.\ref{chi_analysis_5} and Figs. \ref{chi_analysis_60}, \ref{chi_analysis_90} are quite different. When $\theta_0=90^\circ$, there is good symmetry in the contour map which is gradually broken when scattering angle decreases. So long as scattering angle is relatively large ($\theta_0 \ga 20^\circ$), quasi-symmetry is maintained and $\phi_B$ corresponding to the symmetric axes equals to the scattering angle (or to $\theta_0+180^\circ$) (see dash lines in Fig.\ref{chi_analysis_60} and Fig.\ref{chi_analysis_90}). $\chi$ on the quasi-symmetric axes are zero. This means that when magnetic field vector and line of sight (LOS) are coplanar, the vector of polarization also lies in the same plane. Our result also shows that the position angle reduces with the decrease of the scattering angle. Except for very small limited domains (near the parabolic curves), the position angle in Fig.\ref{chi_analysis_5} is generally in the range of $-10^\circ \sim 10^\circ$. When $\theta_0=0^\circ$ (or $180^\circ$), $\chi$ is always $0^\circ$. Therefore, when the scattering angle is small, the direction of polarization generally reflects the magnetic field direction in situ. In addition,, the polarizations are the same for $\theta_0$ and $(180^\circ-\theta_0)$  because of the symmetry. 

\begin{figure}[!ht] 
  \includegraphics[
  width=0.45\textwidth,
  height=0.3\textheight]{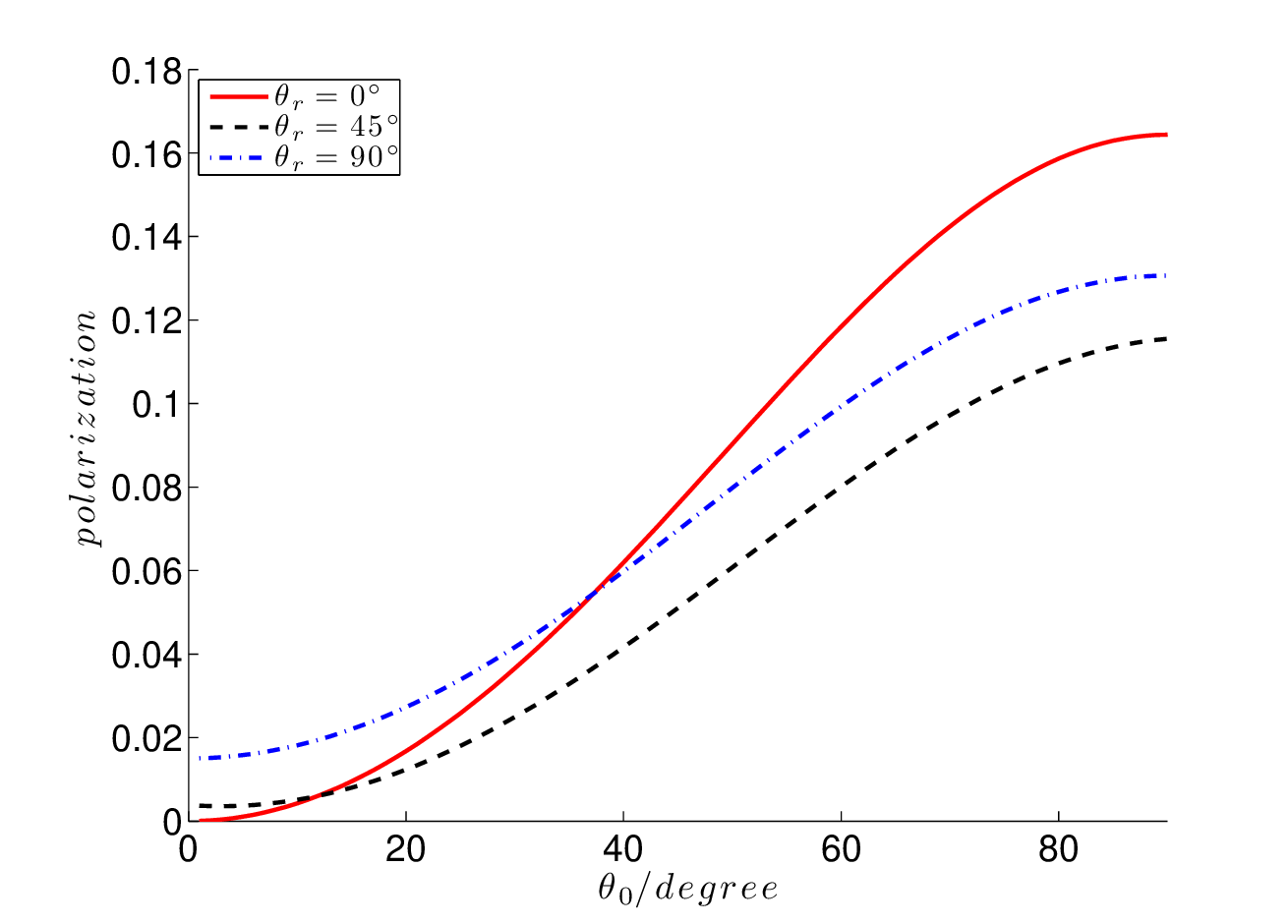}
  \caption{The polarization degree vs. the scattering angle (from $0^{\circ}$ to $90^{\circ}$) in three cases with specific geometry. The red solid line shows the case when the direction of radiation coincides with the magnetic field vector. The polarization degree in this case is the same as that without magnetic field. The black dashed line refers to the case with $\theta_r = 45^\circ$ and line of sight rotate in the plane defined by the radiation and  the magnetic field, therefore, $\theta = 45^\circ - \theta_0$ and $\phi_r = \phi$. The blue dash-dot line represents the case when the magnetic field vector is perpendicular to the scattering plane.}
  \label{var_comp}
\end{figure}

\begin{figure}[!ht] 
  \includegraphics[
  width=0.45\textwidth,
  height=0.32\textheight]{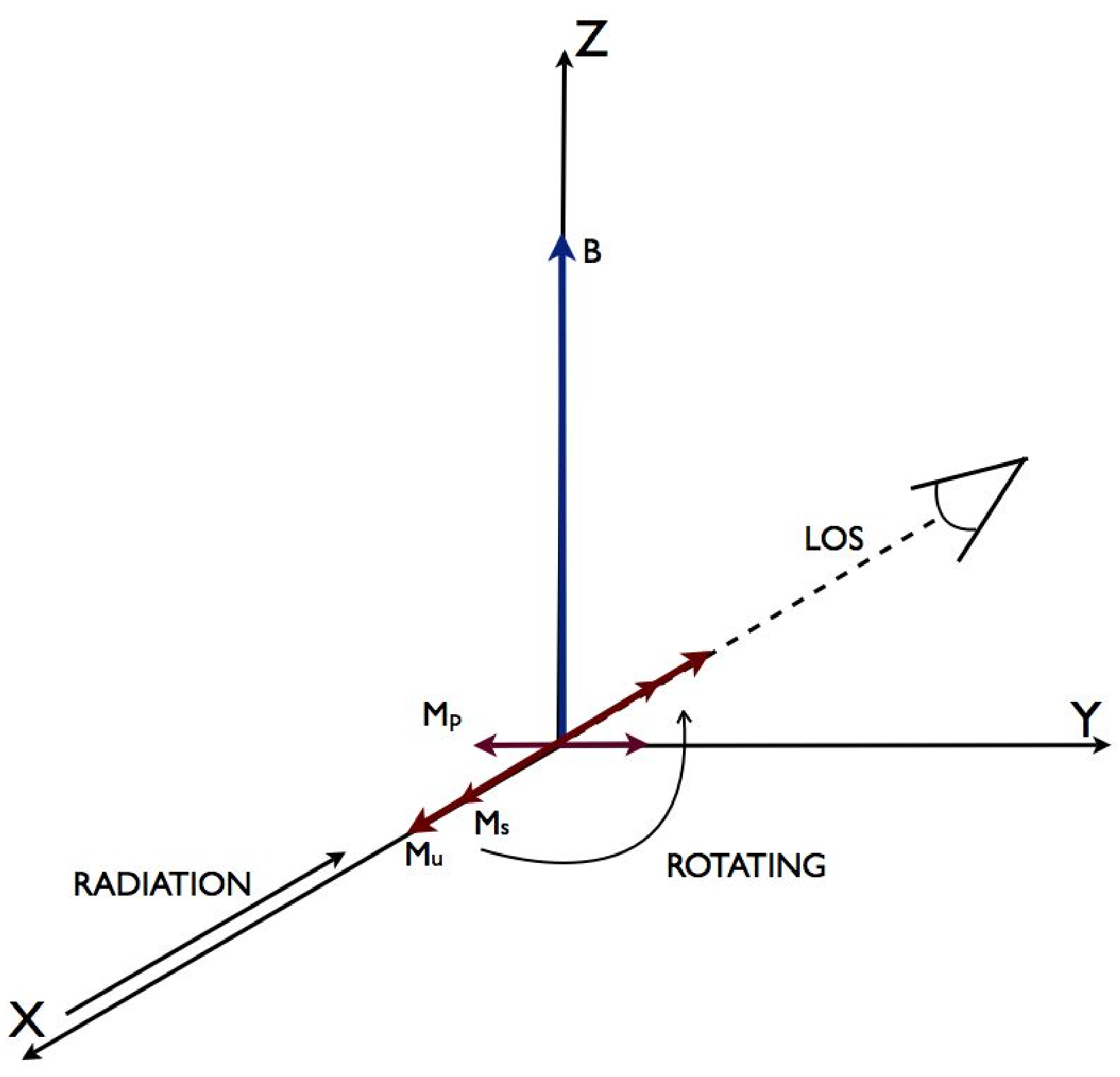}
  \caption{A special case to illustrate how magnetic field enhances polarization degree at small scattering angle. In this case, $\theta_0=180^\circ$, namely,  the radiation and observation directions are both along the x-axis. Magnetic field is along z-axis. Atomic gas is located around the origin. The beam of light will align the atoms, creating a dipole moment, $M_u$, in the same direction. The emission from $M_u$ has zero polarization along the x-axis, the LOS. Thus, if there is no magnetic field, no polarization can be detected. When magnetic field exists, however, $M_u$ is precessing by z-axis, so a part of dipole moment is shared to y-axis, creating $M_p$, which can generate the polarized radiation along LOS.}
  \label{small_scattering}
\end{figure}

When the scattering angle is relatively small ($\lesssim 40^{\circ}$, see Fig.\ref{var_comp}) \footnote{By small scattering angle, we refer to the geometry that line of sight is close to the line of pumping radiation, so that either $\theta_0$ or $\vert 180^\circ - \theta_0 \vert$ $\lesssim 40^{\circ}$ is the small scattering angle case.}, the polarization can increased in the presence of the magnetic field. We plot Fig.\ref{var_comp} to compare with the classical result in \citet{Varshalovich1968}. The lines for $\theta_r=0^\circ$ are almost the same. However, in the case of $\theta_r=45^\circ$, the two figures are different in small $\theta_0$ range. In \citet{Varshalovich1968}, the line of $\theta_r=45^\circ$ line is higher than that of $\theta_r=90^\circ$ when $\theta_0$ is approaching to zero, which is incorrect since they did not account for the coherence in their calculations. If we adopt the semi-classical language by Hanle (see \citealt{Kastler1973}), the GSA created by the anisotropy of the radiation field generates a dipole moment $\vec{M_l}$ (corresponding to ${\rho_l}^2_0$ term of the density matrix of the lower level) along the direction of radiation. Through absorption, the angular momentum is transferred to the upper level,  generating a dipole moment $\vec{M_u}$ (corresponding to ${\rho_u}^2_0$ of upper level), which emits the polarized light. The emission from this dipole moment is axially symmetric, therefore the polarization of the scattered light decreases to zero when LOS is approaching to the direction of radiation. When the magnetic field exists, $\vec{M_l}$, therefore, $\vec{M_u}$ will precess around the magnetic field. The precessing dipole momentum $\vec{M_u}$ can be then decomposed into two parts, $\vec{M_s}$ along the radiation and $\vec{M_p}$ perpendicular to the former. When observing along the direction of radiation (the extreme case for small scattering angle, see Fig.\ref{small_scattering}), $\vec{M_s}$ only emits unpolarized light in LOS for the reason mentioned above. Nevertheless, $\vec{M_p}$ momentum emits polarized light along LOS, so the polarization degree becomes nonzero. This gives us a physical idea how the polarization degree can be enhanced by the magnetic field at small scattering angle. This idea was also employed by \citet{House1974} to explain how the magnetic field decreases the polarization degree at large scattering angle.  In the more accurate quantum description, the existence of magnetic field creates the Zeeman coherence term, ${\rho_l}^2_2$, on the ground level which induces ${\rho_u}^2_2$ on the upper level\footnote{$\rho^2_2$ reflects the coherence of magnetic sublevels}. It is ${\rho_u}^2_2$, generated by the precession of $\vec{M_u}$, that contributes to the enhancement of the polarization of the scattered light. In a nutshell, magnetic field increases the polarization degree at small scattering angle by inducing the Zeeman coherence term ${\rho_u}^2_2$ indirectly.

\begin{figure}[!ht] 
  \includegraphics[
  width=0.45\textwidth,
  height=0.3\textheight]{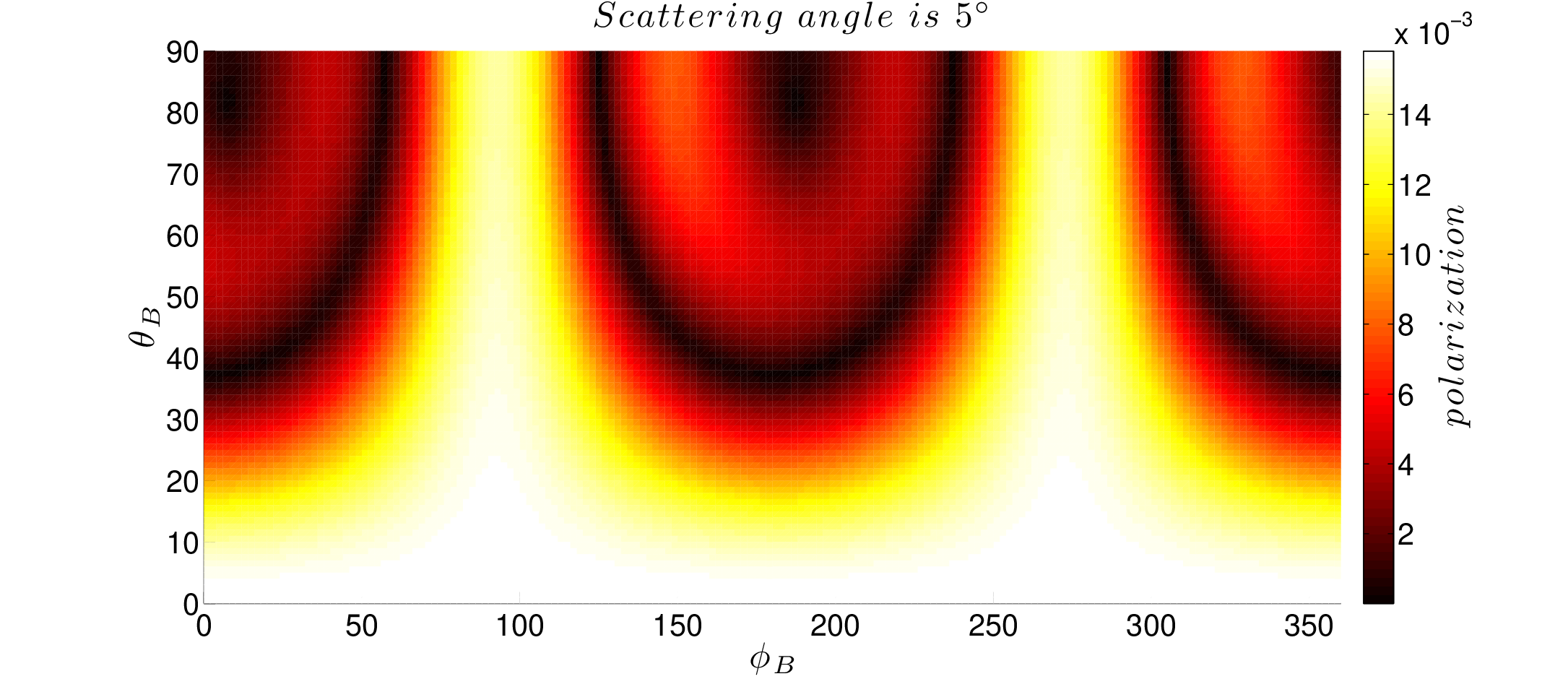}
\caption{The contour map of the polarization degree as a function of the direction of magnetic field. The scattering angle is fixed at $5^{\circ}$. ($\theta_{B}$, $\phi_{B}$) is the solid angle of the magnetic field s defined in Fig.\ref{rotate_dethe}.}
\label{realobs1}
\end {figure}

\begin{figure}[!ht]  
  \includegraphics[
  width=0.45\textwidth,
  height=0.3\textheight]{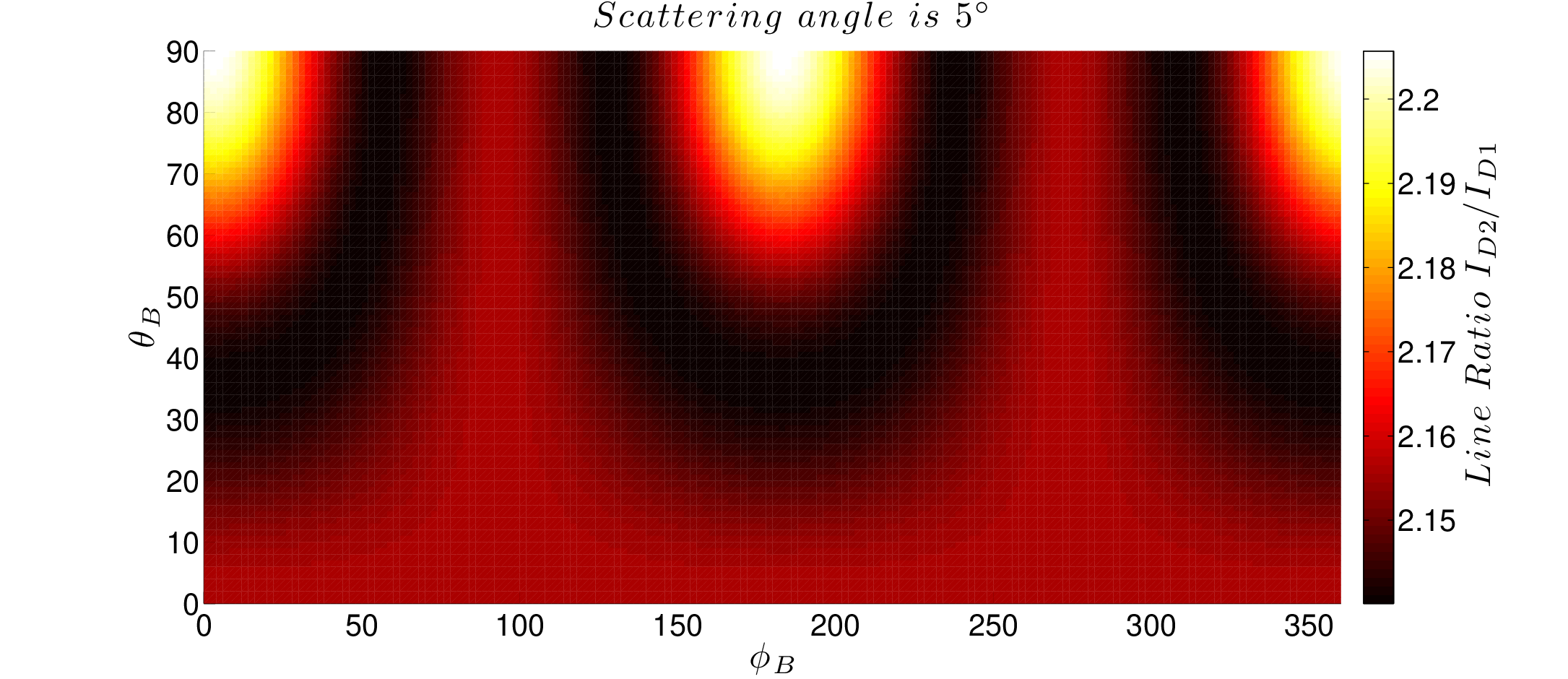}  
\caption{The contour map of line ratio, $I_{D2}/I_{D1}$ as a function of the direction of magnetic field. The scattering angle is also set to be $5^{\circ}$. }
\label{realobs2}
\end {figure}

\section{REAL OBSERVATION}

When it comes to the real observation, the direction of the magnetic field is to be deter-mined according to the results of the spectropolarimetry observations. Suppose we have made observations around Io. Given any fixed scattering angle, we can produce a similar contours maps like Figs.\ref{realobs1} and \ref{realobs2} \footnote{It is more convenient to use the polarization degree and the line ratioto determine the direction of magnetic field, especially, when the scattering angle is large. As discussed in \S4, the directions of magnetic field (projected on the celestial plane) and polarization coincide only when the scattering angle is small.} Then with the observed polarization degree and the line ratio $I_{D2}/I_{D1}$ of each position, we can find the direction of magnetic field ($\theta_{B}, \phi_{B}$) by comparing the results with the contour maps of the polarization degree and line ratio for a fixed $\theta_0$ (see Fig.\ref{realobs1} and Fig.\ref{realobs2}). In general, we can determine the 3D direction of magnetic field with any alignable species using two spectrum lines.

There are some technique issues that one need consider. Both continuum polarization from Halley \citep{Bastien} and  Na D lines from Io have been measured \citep{Brown}. Thanks to the high brightness of the two objects, the observation of the Na D2 line polarization is feasible if the polarization is not too small. The polarization degree is constrained by the Sun-Earth-object geometry. Generally speaking, there is positive correlation between polarization and the scattering angle. In the case of Io and Halley, scattering angles are the phases of the two objects observed from the Earth. As mentioned above, the phase of Halley can be as large as $90^{\circ}$ but the maximum of Io is $11.5^{\circ}$. So one may be concerned about the detectability of the Na D2 line polarization from Io. Our synthetic observation shows that the polarization degree of Io will be generally higher than $1\%$ (see Fig.\ref{syn Io pol}). It can be found from Fig.\ref{realobs1} that there is large domain where t polarization degree is higher than $1\%$. In addition, the low polarization degree can be compensated by the fact that  Io is very bright seen from the Earth. Moreover, the polarization for the small angle scattering can be actually enhanced by the magnetic field. 
\section{DISCUSSION}

Although we focus on the emission line of Na D2, there are abundant atomic species alignable in the diffuse astrophysical environments. In other words, this method can be applied to detect magnetic field in the interplanetary, interstellar and intergalactic space (see \citealt{Yan2012}). It can be also used for the magnetic field near QSOs and other objects, as long as the magnitude of the magnetic field is sub-gauss and the density is not too high.

The resolution of our synthetic observation is limited by the space interval of the magnetic field data that we adopted. The spatial resolution is lower that that of many ground-based telescopes. For instance, the Steward MMT with resolution 0.018 arcsec at the wavelength of 0.5 $\mu$m (see \citealt{Strittmatter}), can resolve a spatial distance less than 100 km\footnote{The resolution of the spectropolarimeter is not considered here.} if the target is near Jupiter. However, the space resolution of Galileo spacecraft, which can be calculated from the data we used, is actually larger than 700km. Hence, the spatial resolution of real observation based on GSA can be higher than that of spacecrafts. In addition,GSA provides us a unique opportunity to detect the magnetic field that is beyond the reach of space probes. 

GSA is sensitive to the orientation of the magnetic field. The polarization direction traces directly the orientation of magnetic field, especially, when the scattering angle is small. We can acquire the entire accurate information of the magnetic field with two or more lines from alignable species. Since alignment happens on a very short timescale (inverse of photo-excitation rate), instantaneous tomography of the magnetic field can be realized through GSA.

\section{SUMMERY}

\begin{enumerate}
  \item Ground State Alignment (GSA) is a powerful and cost-effective method to diagnose small scale magnetic field topology in diffuse medium. It is sensitive to sub-gauss magnetic field, the  common regime of magnetic field in many astrophysical environments, for example, the interplanetary medium, the ISM and the intergalactic medium.
  \item Polarization of scattered light from aligned atoms (or ions) carries the information of local magnetic field.
  \item When scattering angle is small, the polarization degree of scattered light can be enhanced by the magnetic field and the position angle is close to zero, so that the polarization direction traces the direction of magnetic field projected on the celestial sphere.
  \item The two synthetic observations demonstrate that the polarization degree is sufficient for practical spectropolarimetry observations.
 \item Both spatial and temporal variation of magnetic field can be traced by GSA.
\end{enumerate}

\section{APPENDIX}
We obtain Jupiter's magnetic data from NASA PDS-PPI in the frame JSO (Jupiter Solar Orbital) whose illustrations can be found in the corresponding website. \footnote{see \url{http://pds-ppi.igpp.ucla.edu/index.jsp}} In JSO coordinates, X points from Jupiter to the Sun. Z is parallel to the Jupiter orbital plane upward normal. Y completes the right handed set. For the case of comet Halley, the coordinate frame is CSE (CometoCenter Solar Ecliptic), almost the same as JSO except that the center object is the comet. In CSE coordinates, the origin is at comet position, x-axis is from comet to the Sun, z-axis is to ecliptic northpole and (x,y)-plane is parallel to the ecliptic plane.

\acknowledgments

HY and JS are supported by the "985" program of Peking University as well as the ``Be-yond the Horizons" grant from Templeton foundation. HY acknowledge the visiting professorship from the International Institute of Physics (Brazil).

\bibliographystyle{spr-mp-nameyear-cnd}
\bibliography{modified_Io_manuscript}

\end{document}